\documentclass[onecolumn,11pt,a4paper]{IEEEtran}
\usepackage[utf8]{inputenc}
\usepackage[english]{babel}
\usepackage[T1]{fontenc}
\usepackage{array}
\usepackage{graphicx}
\usepackage[caption=false,font=footnotesize]{subfig}
\usepackage{cite}
\usepackage{algorithm}
\usepackage[noend]{algpseudocode}
\usepackage{subfig}
\usepackage{wrapfig}
\usepackage{transparent}
\usepackage[cmex10]{amsmath}
\usepackage{amsfonts,amssymb}
\usepackage{bm}
\usepackage{booktabs}
\usepackage{color}
\usepackage{float}
\usepackage{multicol}
\usepackage{soul}
\usepackage{url}
\usepackage{indentfirst}
\usepackage{lipsum}

\usepackage{xspace,colortbl}

\begin{document}

\title{\bf Multi-Power Level $Q$-Learning Algorithm for Random Access in NOMA mMTC Systems}
\author{{Giovanni~Maciel~Ferreira~Silva}, {Taufik~Abr\~ao}\\
\thanks{G. Maciel and T. Abr\~ao are with the Department of Electrical Engineering, State University of Londrina, Paraná, Brazil. E-mail: giomaciel.fs@gmail.com,\,\, taufik@uel.br}}
\date{\today}

\maketitle

\begin{abstract}
The massive machine-type communications (mMTC) service will be part of new services planned to integrate the fifth generation of wireless communication (B5G). In mMTC, thousands of devices sporadically access available resource blocks on the network. In this scenario, the massive random access (RA) problem arises when two or more devices collide when selecting the same resource block. There are several techniques to deal with this problem. One of them deploys $Q$-learning (QL), in which devices store in their $Q$-table the rewards sent by the central node that indicate the quality of the transmission performed. The device learns the best resource blocks to select and transmit to avoid collisions. We propose a multi-power level QL (MPL-QL) algorithm that uses non-orthogonal multiple access (NOMA) transmit scheme to generate transmission power diversity and allow {accommodate} more than one device in the same time-slot as long as the signal-to-interference-plus-noise ratio (SINR) exceeds a threshold value. The numerical results reveal that the best performance-complexity trade-off is obtained by using a {higher {number of} power levels, typically eight levels}. The proposed MPL-QL {can deliver} better throughput and lower latency compared to other recent QL-based algorithms found in the literature.	
\end{abstract}

%-------------------------------------------------------------------------------
\section{Introduction}\label{sec:intro}
%-------------------------------------------------------------------------------
Machine-type wireless communication will be more widely used in applications such as the internet of things (IoT), smart house, virtual reality, etc. \cite{Chen2020_mMTC,Kalalas2020_mMTC}. The goal of the B5G wireless communications involves achieve ubiquitous communication in networks with ultra-dense device allocation \cite{Chowdhury2020_6G, Nguyen2021_6G_IoT, Lee2021_6G}. A data consumption of nearly five zettabytes per month is estimated across 17 billion devices \cite{Jiang2021_6G}. In addition, due to the outbreak of the COVID-19 pandemic, there has been a remarkable increase in remote activities in work, health and education areas, which will be much more frequent in the post-pandemic environment \cite{Bhat2021_6G}.

Devices connected to the wireless network use different types of service. In the fifth generation of wireless communications (5G) systems, a clear division into three main use modes was defined \cite{Popovski2018}: enhanced mobile broadband (eMBB) for devices that require high data rates as an augmented reality user; ultra-reliable low-latency communications (URLLC) for applications that require 99.999\% communication reliability such as remote surgery, while holding end-to-end latency below 1 ms; and massive machine-type communications (mMTC), composed of thousands of devices with low processing power that access network data sporadically.

The study of these services remains relevant for 6G application scenarios. In the new generation of communications, new services will be generated by merging the benefits of existing ones. In \cite{Jiang2021_6G}, massive ultra-reliable low-latency communication (mULC) is presented as a combination of the low latency of URLLC with the high number of mMTC devices, a densification application process. This new use mode can be associated with intelligent transport, where high reliability is required for traffic safety and various traffic sensors and monitors send data about the vehicle's condition. Besides, ubiquitous mobile broadband (uMBB) is also suggested as a use of high eMBB rates in mMTC devices to enable applications such as ubiquitous networking and digital twin.

As the mMTC scenarios studied in 5G could be associated to the 6G systems, the analysis of the main problems that affect this service is still relevant. One is the random access (RA) procedure. To reduce latency in communication with devices, it is common to use grant-free RA techniques, in which devices do not need a training step with pilot sequences before sending data packets. With the increase in the number of devices and the data rate starvation with new applications, the problem of RA is aggravated. As the device access to the network is sporadic. Still, if there is a crowded number of inactive users in the network, then it is common for two or more devices to select the same resource block to transmit data, characterized as a collision.

Several techniques mitigate the massive RA problem \cite{Jiao2021_RAmMTC,Wang2021_RAmMTC}. One of the simplest and least complex is performed by the slotted ALOHA (SA) protocol, which makes the device resend the collided packet after a fixed time window. There is also the strongest-user collision resolution (SUCRe) protocol \cite{Nishimura2020}, which solves the collision problem by calculating, in a distributed way in each user terminal (UT), the strongest user signal. Another possibility to mitigate the RA issue in (over)-crowded networks is deploying reinforcement learning (RL) techniques. {RL is a trial-and-error-based algorithm with many applications for many known problems in wireless networks. In \cite{Zeng2021_RL_ETT}, an RL-based algorithm is used to increase throughput and solve the problem of collisions between primary and secondary users in a spectrum-sharing environment. In \cite{Huang2021_RL_ETT}, the throughput of a multi-relay system with jamming is increased using the RL algorithm.}

{In RL algorithms applied to the RA problem}, devices take actions and receive rewards from the central node indicating the quality of actions taken. RL has an advantage over more traditional machine learning (ML) techniques in such crowded RA complex scenarios, as it is not necessary to passively receive a dataset \cite{Mohri2018_MLFoundations}.

A more simplified yet effective RL model for this scenario is the $Q$-learning (QL), which is a model-free RL \cite{Otterlo2012_RL}. {Typically, QL algorithms present reduced complexity and they are easy to implement in low-power consumption IoT devices. In \cite{Ge2021_QL_ETT}, QL is used in an IoT environment to increase the success rate of task scheduling. In our elaborated RA scenario, the} device learns which are the best resource blocks it should transmit based on its $Q$-table storage of the rewards sent by the central node. The low complexity of QL makes it suitable to operate in crowded RA scenarios with many devices {randomly} transmitting short packets \cite{Pandey2021_QlearningLPWAN,Tran2021_SCMAQlearning}. In \cite{Sharma2019_Collaborative}, an independent QL technique with a binary reward and a collaborative technique in which the device receives information on the congestion level of each time slot are proposed. In \cite{Valente2020_QLearning_NOMA}, a NOMA-based QL algorithm is proposed in which the device can transmit at up to three different power levels to generate power diversity at the receiver while increasing throughput. {Recently,  \cite{Giovanni2021_P2} proposed a packet-based QL scheme that can benefit} devices that still have many packets to transmit, sending them a bigger reward.

The {\it contribution} of this work is twofold: first, we propose a multi-power levels QL algorithm (MPL-QL), evaluating the impact of increasing power levels on the throughput and latency, differing from what was done in \cite{Valente2020_QLearning_NOMA} where only three power levels are proposed, which does not exploit the full benefit of the power domain of NOMA. Second, we compare performance {metrics, such as throughput,} of the proposed MPL-QL protocol with {four} well-established RA protocols, the SA, the independent QL \cite{Sharma2019_Collaborative}, the collaborative QL \cite{Sharma2019_Collaborative}, {and the packet-based QL \cite{Giovanni2021_P2}}.

The remainder of the {paper} is composed of the system model described in Section \ref{sec:model}; the proposed MPL-QL algorithm is presented in Section \ref{sec:qlearning}; numerical results are analyzed in Section \ref{sec:results}; the final remarks in Section \ref{sec:conclusions} closes the {paper}.

%-------------------------------------------------------------------------------
\section{System model}\label{sec:model}
%-------------------------------------------------------------------------------

\begin{figure}[!htb]
    \centering
    \includegraphics[trim={1cm 10.2cm 1cm 10.5cm},clip,width=0.5\columnwidth]{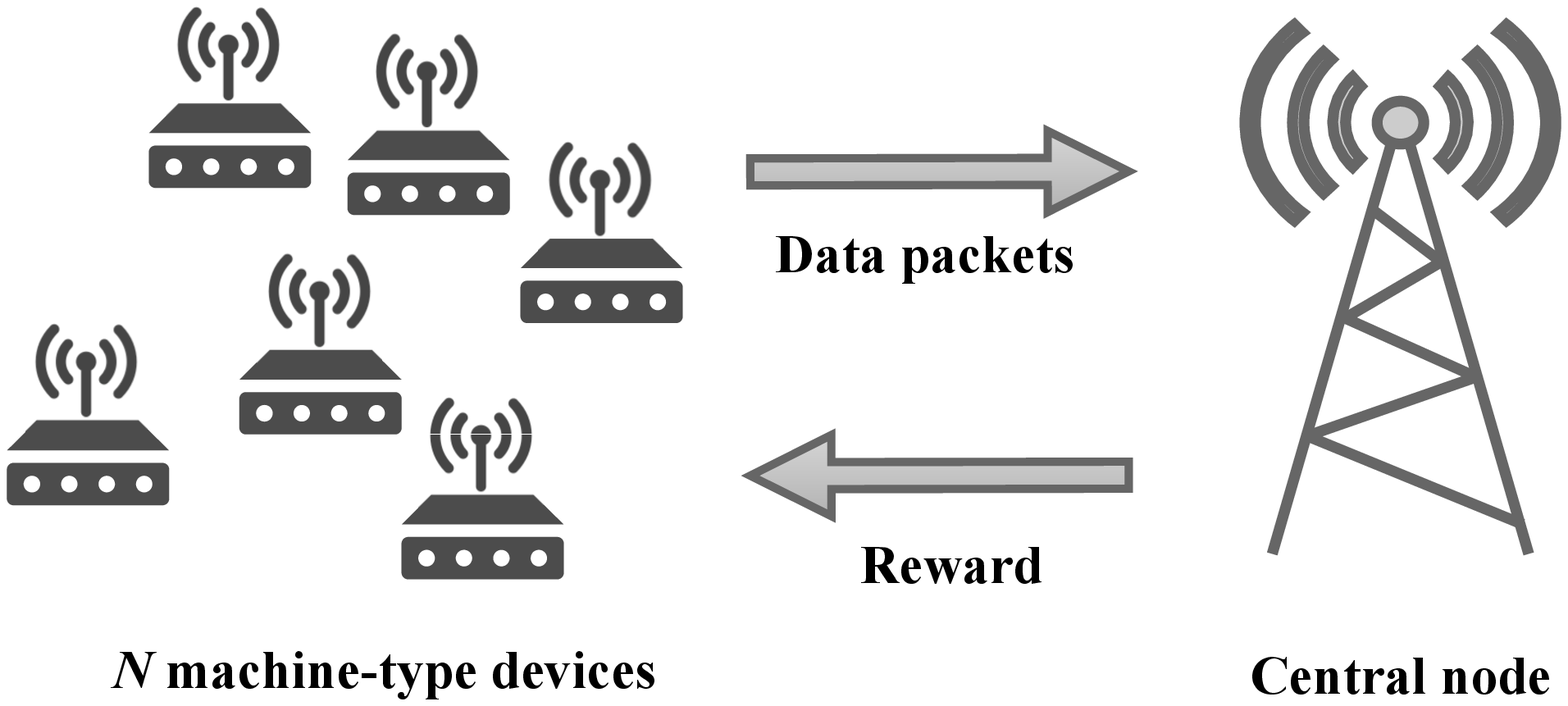}
    \caption{System model.}
    \label{fig:model}
\end{figure}

There are $N$ mMTC devices sending uplink (UL) packets to a central node in a circular cell with radius $r$. 
The frequency resources used are a carrier $f_c$ and a bandwidth $B$. The $n$-th device is $d_n$ meters away from the central node and transmits with power $P_t$. The distribution of devices within the circular cell is shown in Fig. \ref{fig:model}.

\begin{figure}[!htb]
\centering
\includegraphics[clip,width=.6\columnwidth, angle =-90]{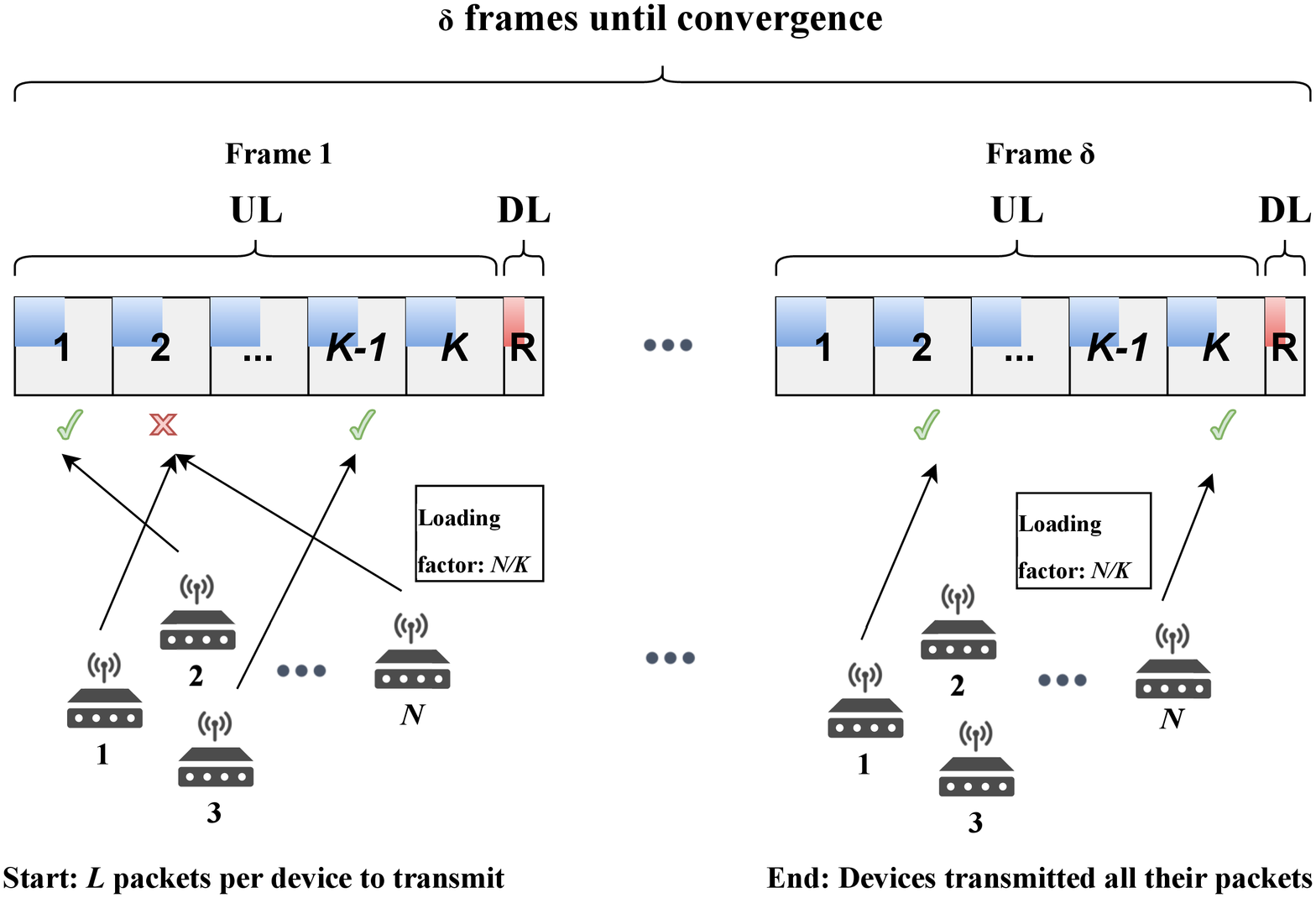}
    \caption{Frame in UL and DL time-slots until the end of system transmission (all devices), namely convergence of transmission process.}
    \label{fig:frame}
\end{figure}

The transmit frame in the UL is divided into $K$ time slots, while a downlink (DL) time-slot at the end is deployed for central node broadcast. The devices randomly select a time-slot to transmit. The set $\psi_k$ contains the indexes of all devices that selected $k$-th time-slot, $k \in \{1,\dots,K\}$. Furthermore, each device has $L$ packets to transmit. The end of system transmission occurs when all devices successfully transmit all of their $L$ packets. At the end, we define the total latency $\delta$ as the total number of spent frames to attain convergence, {\it i.e.}, all packets transmitted successfully by all devices. Assuming that the DL slot is much smaller than the UL slot, it is possible to approximate the length of a frame to $K$ time-slots and the total number of time-slots until the end is $\delta K$. Fig. \ref{fig:frame} shows how transmission frames are divided.

The received signal in the central node at the $k$-th time-slot is simply defined as:
\begin{equation}\label{eq:signal_received}
 y_k = \sum_{\forall n \in \psi_k} x_{n,k} + w_k,
\end{equation}
where $x_{n,k}$ is the attenuated signal transmitted by the $n$-th device at the $k$-th time-slot, and $w_k \sim \mathcal{CN}(0,N_0B)$ is the additive white Gaussian noise (AWGN) at the receiver in the $k$-th time-slot with power spectral density $N_0$.

Let's consider that $h_{n,k}$ is an independent and identically distributed zero mean and unit variance Rayleigh fading of the $n$-th device at $k$-th time-slot. Therefore, the instantaneous signal-to-interference-plus-noise ratio (SINR) received from the $n$-th device at the $k$-th time-slot can be defined as
\begin{equation}\label{eq:sinr}
 \gamma_{n,k} = \dfrac{P_{n,k }}{\sum_{\forall j \in \psi_k, j \neq n}P_{j,k} + w_k^2},
\end{equation}
where $P_{n,k} = h_{n,k}^2\bar{P}_{n}$ is the instantaneous power of the $n$-th device at $k$-th time-slot. $\bar{P}_n$ is calculated based on the log-distance path loss model:
\begin{equation}\label{eq:average_power}
    \bar{P}_{n} = P_t + \bar{P}_{d_0} - 10\eta\log_{10}\left(\dfrac{d_{n}}{d_0} \right), \;\;\; \text{[dB]}
\end{equation}
where $\eta$ is the path loss exponent, $d_0$ is a reference distance, and $\bar{P}_{d_0}$ is a reference constant power given by
\begin{equation}\label{eq:friis}
    \bar{P}_{d_0} = 20\log_{10}\left(\dfrac{c}{4\pi d_0 f_c}\right). \;\;\; \text{[dB]}
\end{equation}
Assuming that the devices have the same quality of service (QoS) requirements, we can set a threshold SINR $\bar{\gamma}$ at the receiver to ensure the packet can be detected. The packet transmitted by the $n$-th device at $k$-th time-slot can be successfully received at the central node when $\gamma_{n,k} \geq \bar{\gamma}$.

%-------------------------------------------------------------------------------
\section{Multi-Power Level $Q$-Learning Algorithm}\label{sec:qlearning}
%-------------------------------------------------------------------------------
This section describes the proposed multi-power level $ Q$-learning-based grant-free RA procedure. Each device can transmit with a maximum power $P_{\max}$. The transmitted symbol is then assumed to have maximum amplitude $V_{\max}$. The symbol $\tilde{x}_n$ transmitted by the device can assume $\mathcal{P}$ equidistant amplitude levels between $-V_{\max}$ and $V_{\max}$, {\it e.g.}, for $\mathcal{P}$ = 4, 
$$
\tilde{x} \in \{-V_{\max},-\dfrac{V_{\max}}{3},\dfrac{V_{\max}}{3},V_{\max}\}.
$$

\begin{figure}[!htb]
\centering
\includegraphics[width=.4\columnwidth]{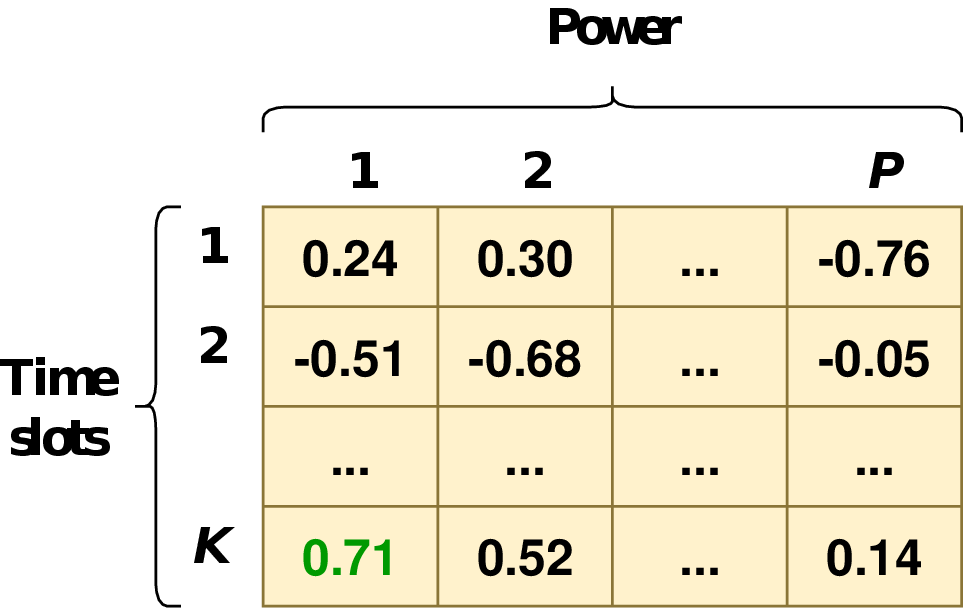}
\caption{$Q$-table for each device.}
\label{fig:qtable}
\end{figure}

The selection of which time-slot and power level the device will transmit is based on the $Q$-table indices whose $Q$-value is maximum. When there are two or more values equal to the maximum, the device randomly selects between them. Fig. \ref{fig:qtable} depicts the structure of the power level and time-slot selection based on the $Q$-table.

As the devices present a power disparity given by the differences in distances and transmission powers, then the central node can apply a {{\it successive interference cancellation} (SIC) procedure} to remove the interference from the devices that collided in the same time-slot. With this, the SINR considering NOMA becomes:
\begin{equation}\label{eq:sinr_noma}
    \gamma_{n,k}^{\textsc{noma}} = \dfrac{P_{n,k}}{\sum_{j = n+1}^{|\psi|}P_{j,k} + w_k^2}.
\end{equation}
The transmission of the $n$-th device is successful if
\begin{equation}\label{eq:reward}
R_{n,k,p} = \begin{cases}
    +1, \;\; \text{if} \;\gamma_{n,k}^{\textsc{noma}} \geq \bar{\gamma}\\
    -1, \;\; \text{otherwise.}
    \end{cases}
\end{equation}
With the reward received, the device updates its $Q$-table \cite{Sutton2018_RL}:
\begin{equation}\label{eq:Q_update}
 Q_{n,k,p}^{(t+1)} = Q_{n,k,p}^{(t)} + \alpha(R_{n,k,p} - Q_{n,k,p}^{(t)}).
\end{equation}
{where $\alpha \in [0,1]$ is the learning rate of the QL algorithm; the closer to 1, the greater the weight that is given to the rewards sent by the central node in relation to the current $Q$-value. If the reward received is $R_{n,k,p}=1$, the success of transmission  decrements the number of packets $\ell_n$ that the $n$-th device still has to transmit.} 

{In the context illustrated in Fig. \ref{fig:frame}, the communication system described in Section \ref{sec:model} requires knowledge of the channel state information (CSI) on both the transmitter and receiver sides to detect the uplink data packets and the downlink reward. As the valuable information is the uplink packets and the reward is binary, then a low-complexity channel estimation technique can be deployed at the device side.}

The devices continue transmitting until all of their $L$ packets are transmitted. Total latency $\delta$ is the number of frames required for the complete transmission of packets until the algorithm converges. Algorithm \ref{algoritmo} indicates the pseudo-code step-by-step of the proposed MPL-QL operation.

\subsection{{MPL-QL Complexity}}

{The complexity of the algorithm can be analyzed in two ways:
\begin{itemize}
    \item {\it Search for the largest $Q$-value within the $Q$-table on the devices side:} increasing $K$ and $\mathcal{P}$ in the system will make the $Q$-table dimension larger, which consequently increases the search space;
    \item {\it Application of the SIC on the central node:} in scenarios where the load factor $\mathcal{L} = N/K$ is very high, {\it e.g.} $\mathcal{L} \geq 5$ in crowded mMTC applications, on average more devices will compete for the available time-slots. The central node needs to apply a SIC to calculate the SINR of each device and check if the transmission was successful or not to send the reward. The higher the $\mathcal{L}$, the greater the complexity and lower the performance in the SIC calculation.
\end{itemize}}

{As there is much more processing power in the central node than in mMTC devices, so the overriding factor of complexity lies in the $Q$-learning algorithm on the device side. Therefore, it is paramount to find a suitable value for the number of power levels $\mathcal{P}$, such that the algorithm performs well without increasing complexity considerably.}

\begin{algorithm}
\caption{\bf MPL-QL algorithm}
\label{algoritmo}
\begin{algorithmic}
\State Initialize $Q_{n,k,p} \sim \mathcal{U}[-1,1]$ $\forall n,k,p$;
		\State Initialize $\ell_n = L, \; \forall n$;
		\State Initialize $\delta = 0, \; S = 0$
		\While{$\sum_{n=1}^{N} \ell_{n} > 0$}
		    \For{all devices that $\ell_n > 0$}
	            \State Search $k$ and $p$ where
	            \State $\; Q_{n,k,p} = \max\limits_{k,p}\{Q_{n,k,p}\}$
	        \EndFor
		    \For{all time-slots $k=1:K$}
		        \If{$|\psi_k| > 0$}
		            \State Calculate $\gamma_{n,k}^{\textsc{noma}}$ using Eq. \eqref{eq:sinr_noma} $\forall n \in \psi_k$
		            \If{$\gamma_{n,k}^{\textsc{noma}} \geq \bar{\gamma}$}
		                \State Success: $S \leftarrow S + 1, \; \ell_n \leftarrow \ell_n - 1$
		                \State $R_{n,k,p}$ = 1
                    \Else
                        \State $R_{n,k,p}$ = -1
		            \EndIf
		            \State Update: $Q_{n,k,p}^{(t+1)} = Q_{n,k,p}^{(t)} + \alpha(R_{n,k,p} - Q_{n,k,p}^{(t)})$
    		    \EndIf
		    \EndFor
		    \State Increment a frame: $\delta \leftarrow \delta + 1$
		\EndWhile
	\end{algorithmic}
\end{algorithm}

%-------------------------------------------------------------------------------
\section{Numerical Results}\label{sec:results}
%-------------------------------------------------------------------------------

This section analyzes the performance and convergence of the MPL-QL algorithm. Performance is measured by throughput and latency, and convergence is analyzed by interference per device and convergence factor.  The system simulations for the QL algorithms were coded in Python language \cite{Habib2019_QLPython}, with Table \ref{tab:parameters} presenting a summary of the parameter values adopted along this section.

\begin{table}[!htbp]
    \centering
    \caption{Numerical parameters.}
    \label{tab:parameters}
    \begin{tabular}{ll}
        \hline
        \textbf{Parameter} & \textbf{Value} \\
        \hline
        Monte-Carlo realizations & $\mathcal{M} = 10000$ \\
        Time-slots per frame & $K$ = 100 \\
        Network loading factor & $\mathcal{L} = \frac{N}{K} \in [0.25;\,\, 10]$ \\
        Packets per device & $L \in $ [50; 100] \\
        Learning rate & $\alpha$ = 0.1\\
        SINR threshold & $\bar{\gamma}$ =  3\\
        Transmit power levels & $\mathcal{P}\in [2; 4; 8; 12; 16]$\\
        \hline
        Cell radius & $r$ = 200 m \\
        Reference distance & $d_0$ = 1 m \\
        Bandwidth & $B$ = 125 kHz \\
        Carrier frequency & $f_c$ = 915 MHz \\
        Path loss exponent & $\eta$ = 3 {(4.77 dB)} \\
        Noise PSD & $N_0$ = $-150$ dBm/Hz \\
        Maximum power & $P_{\max}$ = 1 mW \\
        \hline
    \end{tabular}
\end{table}

{The value of the chosen carrier frequency, bandwidth, and cell size parameters are typical of IoT scenarios, and the amount of devices represents a crowded NOMA mMTC scenario. The typical SINR threshold was selected as $\bar{\gamma}$ = 3, considering that the outage probability is a suitable metric for the communication system performance evaluation \cite{Valente2020_QLearning_NOMA}. Hence, considering the limit of Shannon capacity, if the required SINR is bounded by:
\begin{equation}\label{eq:shannon}
    \gamma_{n,k}^{\textsc{noma}} \geq 2^{\zeta} - 1,
\end{equation}
where $\zeta$ is the spectral efficiency in bits/s/Hz; hence, the adopted (selected) spectral efficiency of $\zeta$ = 2 bits/s/Hz makes $\bar{\gamma}$ = 3.}

\subsection{Throughput and Latency of MPL-QL algorithm} \label{subsec:results_MPL}

\begin{figure}[!htb]
\centering
\includegraphics[trim=7mm 3mm 15mm 12mm, clip=true, width=.6\columnwidth]{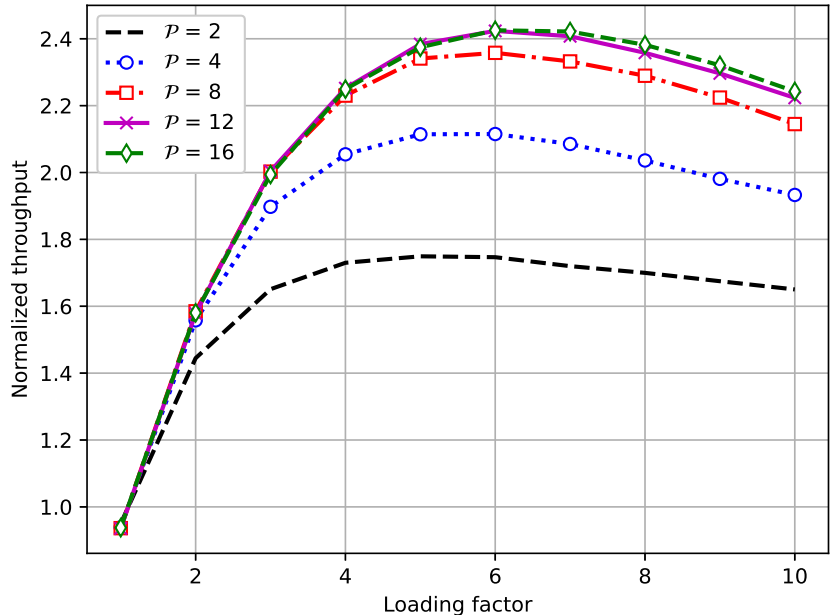}
\caption{MPL-QL throughput for different power levels $\mathcal{P}$.}
\label{fig:throughput}
\end{figure}

Throughput $\tau$ is calculated as the ratio between the total number of successes $S$ and the total number of time-slots required for the algorithm to converge:
\begin{equation}
\tau = \dfrac{S}{\delta K}, \quad \quad \left[\dfrac{\text{success}}{\text{time-slot}}\right]
\end{equation}
Throughput indicates on average how many devices can successfully transmit their packets within a time-slot. In Fig. \ref{fig:throughput}, the throughput of the MPL-QL technique is analyzed as a function of the loading factor for different power levels ($\mathcal{P} \in \{2,4,8,12,16\}$).

Note that the higher the number of power levels, the higher the throughput. Hence, the higher the power levels available at the transmitter, the greater the power difference between the desired signal and the interference at the receiver. This makes the SIC receiver able to detect more packets successfully under a specific limit of the number of power levels since, for a convenient SIC operation, a minimum received power difference between the desired user and the interfering user of the same time-slot must be guaranteed.

When increasing the power levels from 12 to 16, a marginal gain in throughput was observed. In this scenario, there is no significant increase in the power disparity that arrives at the receiver, so a maximum number of possible successes is reached after SIC detection. As the increase in the number of power levels causes an increase in the size of the $Q$-table that the device needs to store, it is possible to say that a good number for power levels is between 8 and 12, as a good performance-complexity trade-off is guaranteed to devices.

\begin{figure}[!htb]
\centering
\includegraphics[trim=6mm 2mm 15mm 12mm, clip=true, width=.6\columnwidth]{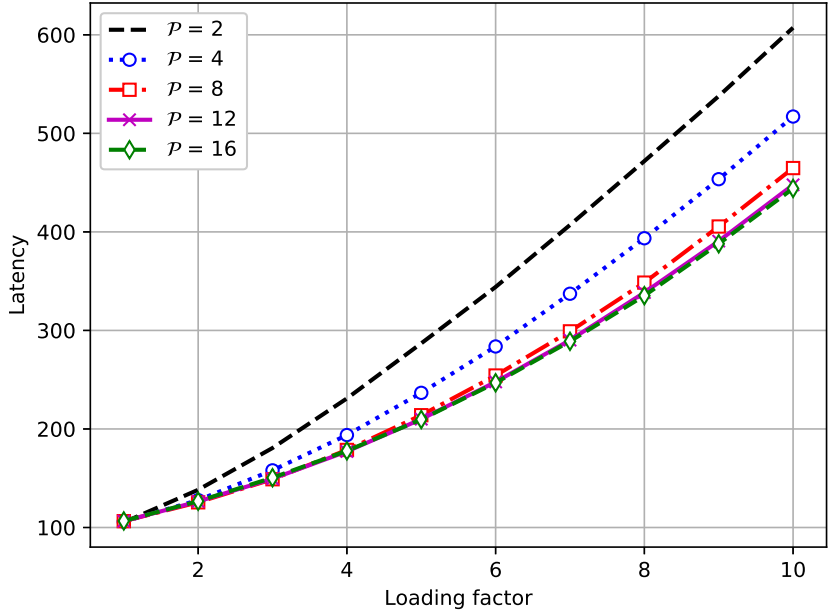}
\caption{MPL-QL latency (total number of frames).}
\label{fig:latency}
\end{figure}

In Fig. \ref{fig:latency}, the latency $\delta$ needed to obtain the algorithm's convergence was analyzed considering the same power levels used in Fig. \ref{fig:throughput}.

Latency decreases with increasing power levels. Considering a loading factor $\mathcal{L}$ = 6, the latency for $\mathcal{P}$ = 8 is $\approx 30$\% lower compared to $\mathcal{P}$ = 2. Analogously to what was discussed in Fig. \ref{fig:throughput}, with a higher value of {number of power levels} $\mathcal{P}$, {the devices have a larger pool of choices of power levels, and they learn from experience which levels are best for transmission, which decreases the probability of collisions and} consequently {decreases} latency.

The differences (reduction) of latency for power levels $\mathcal{P}> 8$ become marginal, which again indicates that a suitable power level guaranteeing a good performance-complexity trade-off is close to 8. This is because the granularity is increased by increasing the power levels. As a result, two or more devices that collide in the same time-slot will have similar power levels. Therefore, the SINR will result higher when the granularity is improved (more power levels), which decreases the number of successes the algorithm can attain. The value of $\mathcal{P} = 8$ power levels was used in the remainder of this section.

\subsection{{Learning Rate for the MPL-QL algorithm}}\label{subsec:learningrate}

\begin{figure}[!htb]
\centering
    \includegraphics[trim=7mm 3mm 15mm 12mm, clip=true, width=.6\columnwidth]{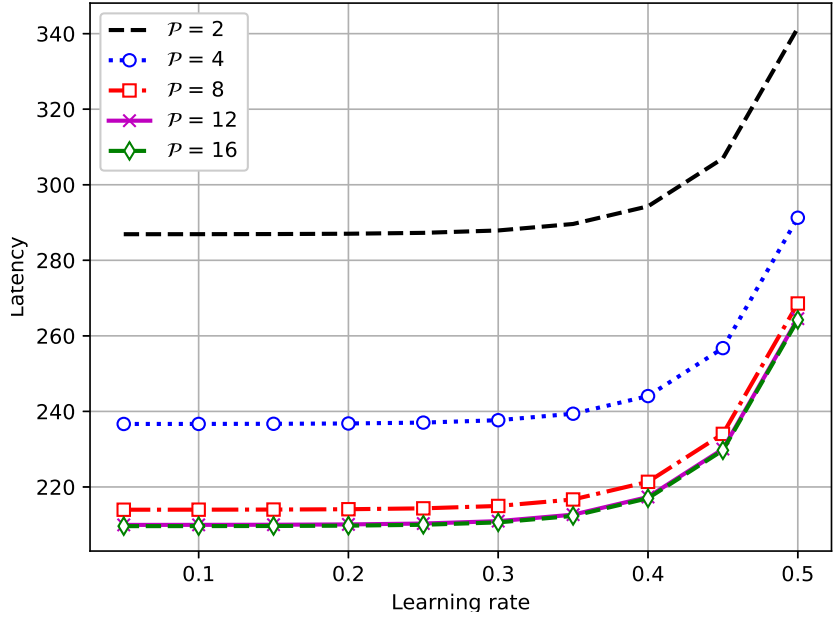}
    \caption{MPL-QL latency for different values of $\alpha$.}
    \label{fig:learningrate}
\end{figure}

{The results of Fig. \ref{fig:throughput} and Fig. \ref{fig:latency} were generated considering $\alpha$ = 0.1, a typical value found in the literature \cite{Sharma2019_Collaborative,Valente2020_QLearning_NOMA,Giovanni2021_P2}, as it considers that each reward sent by the central node weights only 10\% in each $Q$-table update. However, when proposing the MPL-QL algorithm in the NOMA scenario, it is necessary to assess whether the change in the value of $\alpha$ impacts the choice of the most suitable $\mathcal{P}$. Fig. \ref{fig:learningrate} shows the latency of the MPL-QL algorithm when the learning rate varies within a range $\alpha \in [0.05; 0.5]$. In this numerical result, $K$ = 100 time-slots and a loading factor $\mathcal{L}$ = 5 have been considered.}

{The MPL-QL algorithm presents an approximate constant latency up to $\alpha$ = 0.3 because the weight given to the rewards is low. $Q$-table changes are made smoothly so that the wrong decisions at the devices regarding the transmission resources do not negatively impact the packet deliver evolution of the proposed algorithm.}

{The algorithm latency grows substantially by considering increasing learning rate scenarios in the $0.3\leq \alpha < 0.5$. As $\mathcal{L} = 5$ is a very crowded load factor, it is natural that many collisions occur and many negative rewards are sent by the central node since there are, on average, five devices disputing the same time-slot. As $\alpha$ increases, the weight given to negative rewards grows. Therefore, devices make more wrong decisions and take longer to find the best time and power resources to transmit packets successfully.}

{On the other hand, as the value of $\mathcal{P}$ increases, the delay in delivering the total number of packets of all users decreases. However, from $\mathcal{P}$ = 8, the decrease in latency becomes marginal, corroborating the same behavior already observed in the results of throughput (Fig. \ref{fig:throughput}) and latency {\it vs.} loading factor in Fig. \ref{fig:latency}. Hence, $\mathcal{P}$ = 8 represents a suitable performance-complexity trade-off for the proposed MPL-QL algorithm. Finally, from the numerical experiments in Fig. \ref{fig:throughput}, \ref{fig:latency}, and \ref{fig:learningrate}, one can infer that changes in the learning rate do not dramatically affect the choice of the most suitable value for the number of power levels $\mathcal{P}$.}

\subsection{Convergence of MPL-QL algorithm}

The convergence of the MPL-QL algorithm is analyzed by two figures of merit: interference per device and convergence factor per device. The analysis was performed only for the $n$-th device, but on average, the figures of merit for all devices reveal the same behavior.

\begin{figure}[!htb]
\centering
\includegraphics[trim=3mm 5mm 3mm 4mm, clip=true, width=.6\columnwidth]{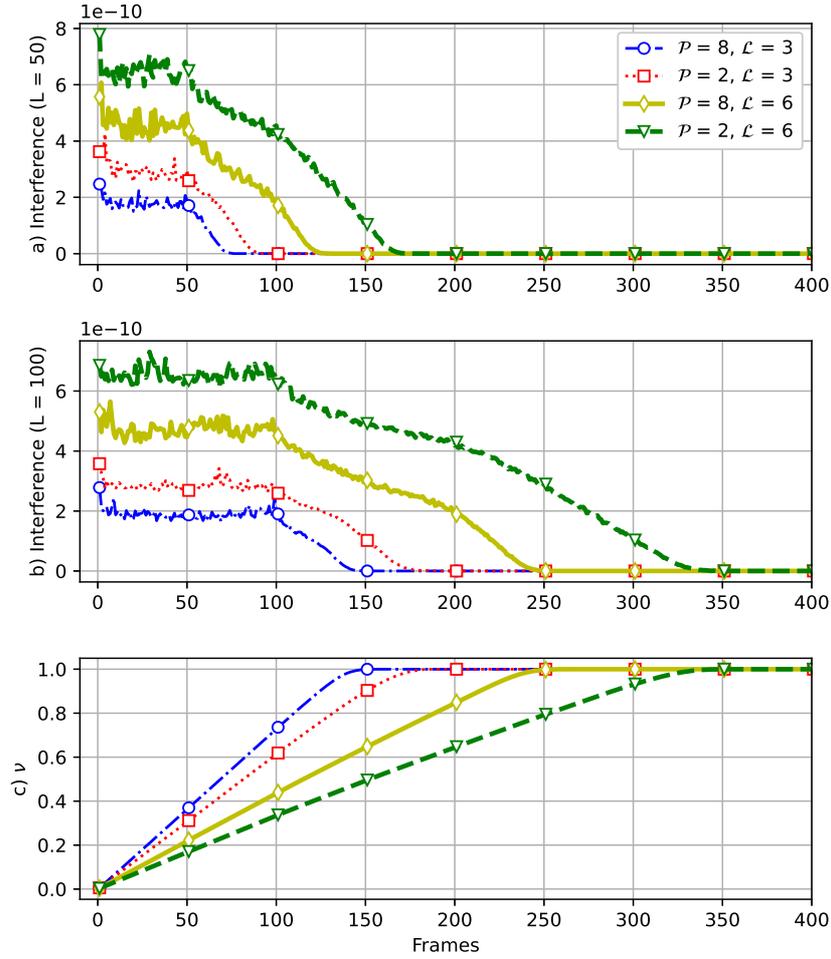}
\caption{Interference and convergence factor of the MPL-QL algorithm considering the $n$-th device. a) $L$ = 50 packets; b) $L$ = 100 packets; c) convergence factor under $L$ = 100 packets.}
\label{fig:interference}
\end{figure}

The interference $I_{n,k}$ of the $n$-th device at $k$-th time-slot is calculated as the sum of the powers of the interfering devices that selected the same time-slot, defined by the subset $\psi_k$;  after SIC detection such interference can be calculated as
\begin{equation}
    I_{n,k} = \sum_{j = n+1}^{|\psi_k|}P_{j,k}, \quad \quad [W]
\end{equation}
while the convergence factor is defined as
\begin{equation}\label{eq:conv_factor}
    \nu_n = \dfrac{L - \ell_n}{L}.
\end{equation}
At the beginning of the algorithm execution, the $n$-th device transmitted only a few packets successfully, so $\nu_n \rightarrow 0$. When the algorithm is close to the convergence, the $n$-th device has already transmitted most of its packets, making $\nu_n \rightarrow 1$. Fig. \ref{fig:interference} depicts the interference along the frames for the $n$-th device considering a) $L=50$, b) $L=100$ [packets/device], and c) convergence factor {\it vs.} the frame ($\delta$) evolution until convergence ($\nu_n=1$).

At the beginning of transmission frames, the MPL-QL algorithm can make wrong decisions in choosing the best time-slots and power levels to transmit. Hence, there is an oscillating behavior of high interference in the early frames. After a latency $\delta\approx L$ frames,  {\it i.e.}, $L = 50$ and  $100$ frames in Fig. \ref{fig:interference}, device interference starts to decrease steadily as devices have already passed the initial learning phase and begin to discover better time-slots and power levels to transmit their packets with a greater probability of success. Indeed, when the number of frames equals $L$, many devices have already transmitted all their packets, as they initially selected the least congested time-slots. Therefore, more empty time-slots start to appear for the $n$-th device, which makes their interference monotonically decrease after $L$ frames.

Increasing the loading factor causes more devices to collide in the same time-slot, which causes an increase in the average interference per device. For this reason, in Fig. \ref{fig:interference}a) and \ref{fig:interference}b), the two scenarios with a loading factor $\mathcal{L}$ = 6 have more significant interference compared to $\mathcal{L}$ = 3.
Moreover, increasing power levels makes the average interference lower. With more power levels, the greater the signal power disparity of the devices that collide in the same time-slot, making the difference between the power of interest and the interfering one greater.

By increasing the loading factor $\mathcal{L}$, more devices are transmitting in the available time-slots. This causes the interference to increase, causing more collisions to happen and increasing the convergence time of the algorithm. For this reason, it is possible to observe that the curves for $\mathcal{L}$ = 3 converge faster than the curves for $\mathcal{L}$ = 6.

Increasing the number of power levels also makes convergence occur faster. This is because more available power levels generate a power disparity between two or more devices that collide in the same time-slot, making the signal from the device with the highest power level even more remarkable in relation to the interfering ones. In Fig. \ref{fig:interference}c), it can be seen that the number of power levels $\mathcal{P}$ = 8 converged faster than the $\mathcal{P}$ = 2 in both loading factor scenarios.

\subsection{Comparison with Other RA Methods}\label{subsec:comparison}

The performance of the proposed MPL-QL algorithm was compared with other methods available in the literature, specifically: \textbf{\textit{a}}) Slotted Aloha (SA), where there is no feedback from the central node to the devices. The devices only send all their UL packets and the number of successes is obtained when there is no collision;
\textbf{\textit{b}}) Independent QL \cite{Sharma2019_Collaborative}; \textbf{\textit{c}}) Collaborative QL \cite{Sharma2019_Collaborative}; and Packet-Based QL \cite{Giovanni2021_P2}.

{All QL algorithms are applied to a NOMA scenario, as there is no single orthogonal time-slot allocation for each device, as devices randomly select which time-slot to transmit. The central node uses SIC to eliminate interference between devices, so the SINR is calculated using Eq. \ref{eq:sinr_noma} to decide if the transmission was successful.}

 As the QL mentioned above techniques do not consider different transmitter power levels, the transmitted power is the same for all devices. This impacts the evolution of the QL algorithm{; hence, as the $Q$-table reveals in Fig. \ref{fig:qtable}, it} does not present the dimension of the powers for such techniques, being considered only the time-slots dimension. Thus, the device learning process is performed only to find the best time-slot for transmission with minimal {probability} of collision.

The difference between the three QL-based algorithms deployed in the comparison is how the central node does the reward. Hence, in the {\it independent QL} algorithm, the reward sent by the central node is {defined as}:
\begin{equation}\label{eq:reward_ind}
R_{n,k}^{\textsc{ind}} = \begin{cases}
+1, \;\; \text{if} \;\gamma_{n,k}^{\textsc{noma}} \geq \bar{\gamma}\\
-1, \;\; \text{otherwise.}
\end{cases}
\end{equation}
It is a binary reward, similar to the MPL-QL, but it is performed only in the time-slot dimension. On the other hand, for the collaborative QL, the congestion level of the time-slot is defined as:
\begin{equation}
    C_k = \dfrac{|\psi_k|}{N};
\end{equation}
and included in the adverse reward of {\it collaborative QL}:
\begin{equation}\label{eq:reward_col}
R_{n,k}^{\textsc{col}} = \begin{cases}
 +1, \;\; \text{if} \;\gamma_{n,k}^{\textsc{noma}} \geq \bar{\gamma}\\
-C_k, \;\; \text{otherwise.}
\end{cases}
\end{equation}

As a result, the collaborative QL algorithm is more complex than independent QL since the central node needs to know the number of devices colliding in each time-slot. However, the performance is superior as more information related to the system state is sent during the execution of the algorithm \cite{Sharma2019_Collaborative}.

The {\it packet-based QL} considers the convergence factor of Eq. \eqref{eq:conv_factor} and includes {such factor in the its} reward:
\begin{equation}\label{eq:reward_pac}
R_{n,k}^{\textsc{pac}} = \begin{cases}
    +1, \hspace{11mm} \text{if} \;\gamma_{n,k}^{\textsc{noma}} \geq \bar{\gamma}\\
-{\dfrac{L - \ell_n}{L}}, \;\; \text{otherwise.}
    \end{cases}
\end{equation}

Packet-based QL random access strategy favors devices that still have a lot of packets to transmit, sending them a greater reward {w.r.t.} devices that are already close to convergence \cite{Giovanni2021_P2}.

\begin{figure}[!htbp]
\centering
\includegraphics[trim=2mm 4mm 3mm 3mm, clip=true,width=.6\columnwidth]{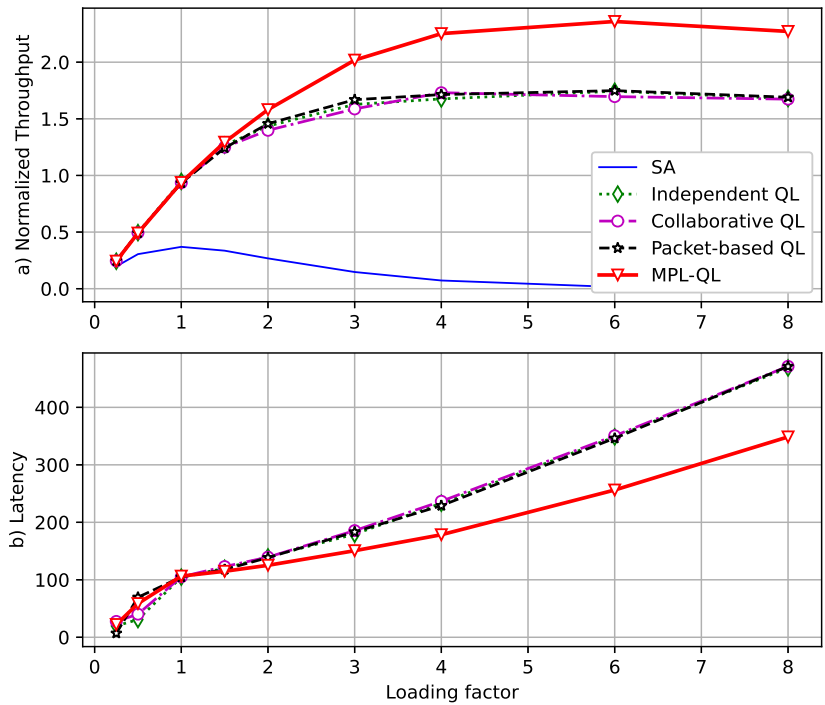}
\caption{a) Throughput, and b) Latency for the SA and four QL-based algorithms, with $\mathcal{P}$ = 8 for the proposed MPL-QL. $L$ = 100 and $K$ = 100.}
\label{fig:comparison}
\end{figure}

Fig. \ref{fig:comparison}.a) shows the normalized throughput and Fig. \ref{fig:comparison}.b) {depicts the latency against loading factor $\mathcal{L}$ for the proposed MPL-QL algorithm, the Slotted Aloha, and the other three QL-based algorithms in the literature. For these results,} $K$ = 100 time-slots/frame and $L$ = 100 packets per device were considered. SA is the most straightforward RA protocol since devices randomly select a time-slot with no reward sent by the central node to indicate transmission quality. For this reason, the SA throughput is the worst among all the analyzed techniques. Independent and collaborative QL techniques have higher throughput than SA as central node rewards are used for devices to better select which time-slots to transmit. In \cite{Sharma2019_Collaborative}, it is noted that the collaborative QL performs better than the independent QL. However, by adding the power domain in NOMA scenarios, the performance of the techniques becomes the same. 

The proposed MPL-QL {random access method has presented a substantial increase in the throughput and simultaneously decrease in the latency for higher loading factors, $\mathcal{L}$. Such system throughput and latency improvements can be explained by the fact that} in the realistic scenario where devices are subject to the effects of path loss and fading, increasing the power diversity at the transmitter allows nearby devices to transmit at different powers, which increases the SINR after the SIC, at the receiver side. Therefore, more successes are expected when using MPL-QL compared to other QL algorithms, increasing the throughput.

{Elaborating further, when analyzing the throughput in Fig. \ref{fig:comparison}a) and latency in Fig. \ref{fig:comparison}b), one infers that the} power domain can be advantageous to allocate more devices in a time-slot. The QL{-based} algorithms in the literature exploring only the time-slot domain do not take advantage of the power diversity in the transmitter to avoid collisions, so they tend to converge more slowly. {Hence, for} a loading $\mathcal{L} = {4}$, where there are on average {4} devices transmitting per time-slot, the SA protocol is capable of generating little more than {$\frac{1}{10}$} success. On the other hand, independent and collaborative QL-based RA algorithms are able to generate $1.7$ successes, while the proposed MPL-QL generates $2.3$ successes. Therefore, it is shown that, on average, the MPL-QL is able to better deal with the collisions between devices, mainly when the loading factor increasing beyond $3$.

%---------------------------------
\subsection{{QL-based RA Techniques with Imperfect SIC}}
%---------------------------------
The previous results were obtained considering that the central node applies a perfect SIC when receiving packets. However, error-free cancellation is difficult to achieve in crowded mMTC scenarios {due to the different levels of interference affecting each signal device}. In this subsection, we consider an imperfect SIC model in which there is a residue of the powers of devices that have already passed through the SIC {modeling the poor signal-canceling effect. Hence, considering NOMA, the new SINR with the imperfect SIC can be written as:}
\begin{equation}\label{eq:sinr_imperfect}
    \tilde{\gamma}_{n,k}^{\textsc{noma}} = \dfrac{P_{n,k}}{\beta\sum_{j = 0}^{n-1}P_{j,k} + \sum_{j = n+1}^{|\psi|}P_{j,k} + w_k^2}.
\end{equation}
where $\beta \in [0,1]$ is the  SIC error factor. $\beta = 0$ indicates that the interference is perfectly cancelled{, collapsing in Eq. \eqref{eq:sinr_noma}, while when $\beta = 1$, models the absence of SIC procedure} at the central node. Typical realistic values for the error factor are in range {$\beta\in \{0.01;\,\, 0.30\}$}, depending on the level of interference and the received power disparities distribution.

\begin{figure}[!htbp]
\centering
\includegraphics[width=.9\columnwidth]{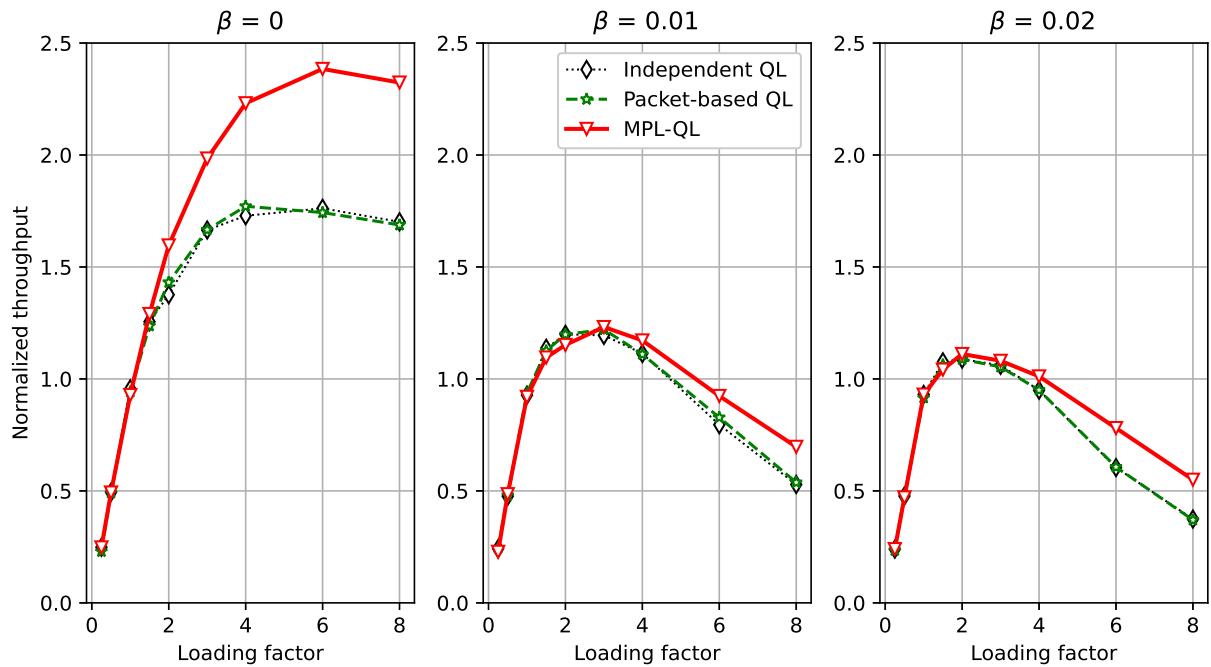}
\caption{Independent QL, Packet-based QL and MPL-QL under SIC imperfection: a) $\beta$ = 0; b) $\beta$ = 0.01; $\beta$ = 0.02. We considered $L$ = 100 and $K$ = 100.}
\label{fig:sic_error}
\end{figure}

Fig. \ref{fig:sic_error} depicts a comparison of the throughput of QL-based techniques considering $\beta \in [0; \, 0.1; \, 0.2]$. As expected, increasing $\beta$ worsens the throughput of all algorithms as interference increases, increasing collision and latency until the algorithm attains convergence. Notice that increasing the value of $\beta$ decreases the maximum number of devices the system serves. MPL-QL achieves maximum throughput for a loading factor $\mathcal{L}$ = 6 {operating under perfect SIC, {\it i.e.,}} $\beta$ = 0. For $\beta$ = .01 and $\beta$ = 0.02, the loading factor to achieve maximum throughput {is reduced to $\mathcal{L} = 3$ and $\mathcal{L} = 2$, respectively}. Such a decrease is also due to increased latency and interference that can not be canceled ($\beta\neq0$). The MPL-QL {method} proved to be superior to the other QL-based algorithms in all {RA crowded} scenarios, {thanks to the greater power-level granularity, allowing the power differences} between the desired device and the interferers larger, reducing collisions.

%-------------------------------------------------------------------------------
\section{Conclusions}\label{sec:conclusions}
%-------------------------------------------------------------------------------
The performance and convergence of the proposed MPL-QL method for different {power-levels granularity} have been characterized {and compared with other QL-based RA algorithms}. It was observed that the best number of power levels that guarantee a good performance-complexity trade-off is $\mathcal{P} = 8$ levels. This value was used to compare the throughput with other recent grant-free RA algorithms, {namely the independent, collaborative, and {packet-based} QL-based algorithms and the classical SA method.} The $8$-levels MPL-QL technique has revealed the best performance compared to the other analyzed RA techniques due to the {enough} power diversity generated by the MPL-QL technique, improving the SINR at the receiver side while increasing the chance of successful transmissions {of a more significant number of devices in crowded RA scenarios}.

{The proposed MPL-QL {method} demonstrated superiority in both throughput and latency regarding the other QL-based algorithms in all RA crowded scenarios analyzed, due to the greater power-level granularity, allowing the power differences between the desired device and the interferers to be larger, reducing collisions.}

 \section*{Acknowledgment}
This work was supported in part by the Coordenação de Aperfeiçoamento de Pessoal de Nível Superior - Brazil (CAPES) - Finance Code
001, in part by the National Council for Scientific and Technological
Development (CNPq) of Brazil under Grant 310681/2019-7, and in part by Fundação Araucária, Paraná State, Brazil under Grant PBA-2016.
%------------------------------------------------------------------------------

\end{document}